\documentclass[showpacs,graphicx]{revtex4}
\usepackage{graphicx}
\usepackage{color}
\begin{document}

\title{Feasible logic Bell-state analysis with linear optics}

\author{Lan Zhou,$^{1,2}$ and Yu-Bo Sheng$^{2\ast}$}
\address{$^1$ College of Mathematics \& Physics, Nanjing University of Posts and Telecommunications, Nanjing,
210003, China\\
 $^2$Key Lab of Broadband Wireless Communication and Sensor Network
 Technology,Nanjing University of Posts and Telecommunications, Ministry of
 Education, Nanjing, 210003,
 China\\
\footnote{Email address: shengyb@njupt.edu.cn}}

\begin{abstract}
\textbf{Abstract}: We describe a feasible logic Bell-state analysis protocol by employing the logic entanglement to be the robust concatenated Greenberger-Horne-Zeilinger (C-GHZ) state. This protocol only uses polarization beam splitters and half-wave plates, which are available in current experimental technology. We can conveniently identify two of the logic Bell states. This protocol can be easily generalized to the arbitrary C-GHZ state analysis. We can also distinguish two $N$-logic-qubit C-GHZ states. As the previous theory and experiment both  showed that the C-GHZ state has the robustness feature, this logic Bell-state analysis and C-GHZ state analysis may be essential for linear-optical quantum computation protocols whose building blocks are logic-qubit entangled state.
\end{abstract}

\maketitle

\section*{Introduction}
Quantum entanglement is of vice importance in future quantum communications, quantum computation and some other quantum information processing procotols \cite{teleportation,Ekert91,QSS,QSDC1,QSDC2}. For example, quantum teleportation \cite{teleportation}, quantum key distribution (QKD) \cite{Ekert91}, quantum secret sharing (QSS) \cite{QSS}, quantum secure direct communication (QSDC) \cite{QSDC1,QSDC2} and quantum repeater \cite{repeater1,repeater2} all require the entanglement to set up the quantum channel.
For an optical system, the photonic entanglement is usually encoded in the polarization degree of freedom. Besides the polarization entanglement, there are some other types of
entanglement, such as the hybrid entanglement \cite{hybrid1,hybrid2,hybrid3,hybrid4}, in which the entanglement is between different degrees of
freedom of a photon pair.  The photon pair can also entangle in more than one degree of freedom, which is called the hyperentanglement \cite{hyper1,hyper2,hyper3,hyper4,hyper5}. Both the hybrid entanglement and the hyperentanglement have been widely used in quantum information processing \cite{hybridusing5,bell8,bell9,bell13,bell14,hyper6}.

Different from the entanglement encoded in the physical qubit directly, logic-qubit entanglement  encodes the single
physical quantum state which
contains many physical qubits in a logic quantum qubit.  Logic-qubit entanglement has been discussed in both theory and experiment. In 2011, Fr\"{o}wis and D\"{u}r described a new kind of logic-qubit entanglement, which shows similar features as the Greenberger-Horne-Zeilinger (GHZ) state \cite{cghz1}.
This logic-qubit entangled state is named the concatenated GHZ (C-GHZ) state. It is also called the macroscopic Schr\"{o}dinger's cat superposed state \cite{cghz2,cghz3,cghz4,cghz5}.
The C-GHZ state can be written as
\begin{eqnarray}
|\Phi_{1}^{\pm}\rangle_{N,M}=\frac{1}{\sqrt{2}}(|GHZ^{+}_{M}\rangle^{\otimes N} \pm |GHZ^{-}_{M}\rangle^{\otimes N}).\label{logic}
\end{eqnarray}
Here, $N$ is the number of logic qubit and $M$ is the number of physical qubit in each logic qubit, respectively.
States $|GHZ^{\pm}_{M}\rangle$ are the standard $M$-photon polarized GHZ states as
\begin{eqnarray}
|GHZ^{\pm}_{M}\rangle=\frac{1}{\sqrt{2}}(|H\rangle^{\otimes M}\pm|V\rangle^{\otimes M}),
\end{eqnarray}
 where $|H\rangle$ is the horizonal polarized photon and  $|V\rangle$ is the vertical polarized photon, respectively.  Fr\"{o}wis and D\"{u}r revealed that the C-GHZ state has its natural feature to immune to the noise \cite{cghz1}. Recently, He \emph{et al.} demonstrated the first experiment to prepare the C-GHZ state \cite{cghz6}.
In their experiment, they prepared a C-GHZ state with $M=2$ and $N=3$ in an optical system. They also investigated
the robustness feature of C-GHZ state  under different noisy models. Their experiment verified that the C-GHZ state can tolerate more  bit-flip and phase shift noise
 than polarized GHZ state. It shows that the C-GHZ state is useful for large-scale fibre-based quantum networks and
multipartite QKD schemes, such as QSS schemes and third-man quantum cryptography \cite{cghz6}.

  On the other hand, similar to the importance of the controlled-not (CNOT) gate to the standard quantum computation model, Bell-state analysis plays the key role in the quantum communication. The main quantum communication branches such as quantum teleportation,  QSDC all require the Bell-state analysis. The standard Bell-state analysis protocol, which utilizes linear optical elements and single-photon measurement can unambiguously discriminate two Bell-states among all four orthogonal
 Bell states \cite{bellstateanalysis1,bellstateanalysis2,bellstateanalysis3}. By exploiting the ancillary states or hyperentanglement,  four polarized Bell states can be improved or be completely distinguished \cite{bell8,bellstate4,bellstate5}. For example, with the help of spatial modes entanglement, Walborn \emph{et al.} described an important approach to realize the polarization Bell-state analysis \cite{bell8}. The Bell-state analysis for hyperentanglement were also discussed \cite{bell13,bellstate6,bellstate7,bellstate8}. By employing a logic qubit in GHZ state, Lee \emph{et al.} described the Bell-state analysis for the logic-qubit entanglement  \cite{logicbell1}.  The logic Bell-state analysis with the help of CNOT gate, cross-Kerr nonlinearity and photonic Faraday rotation were also described \cite{logicbell2,logicbell3,logicbell4}. Such protocols which based on CNOT gate, cross-Kerr nonlinearity and photonic Faraday rotation are hard to realize
 in current experiment condition.

In this paper, we will propose a feasible protocol of logic Bell-state analysis, using only linear optical elements, such as polarization beam splitter (PBS) and half-wave plate (HWP).
Analogy with the polarized Bell-state analysis, we can unambiguously distinguish two of the four logic Bell states.
This approach can be easily generalized to the arbitrary C-GHZ state analysis. We can also identify two of the $N$-logic-qubit C-GHZ states.
As the logic-qubit entanglement is more robust than the polarized GHZ state, this protocol may provide a competitive approach in  future quantum information processing.

\section*{Results}
 The basic principle of our protocol is shown in Fig. 1. The four logic Bell states can be described as
\begin{eqnarray}
|\Phi^{\pm}\rangle_{AB}=\frac{1}{\sqrt{2}}(|\phi^{+}\rangle_{A}|\phi^{+}\rangle_{B}\pm|\phi^{-}\rangle_{A}|\phi^{-}\rangle_{B}),\nonumber\\
|\Psi^{\pm}\rangle_{AB}=\frac{1}{\sqrt{2}}(|\phi^{+}\rangle_{A}|\phi^{-}\rangle_{B}\pm|\phi^{-}\rangle_{A}|\phi^{+}\rangle_{B}).\label{bell1}
\end{eqnarray}
Here, $|\phi^{\pm}\rangle$ and $|\psi^{\pm}\rangle$ are four polarized Bell states of the form
\begin{eqnarray}
|\phi^{\pm}\rangle=\frac{1}{\sqrt{2}}(|H\rangle|H\rangle\pm|V\rangle|V\rangle),\nonumber\\
|\psi^{\pm}\rangle=\frac{1}{\sqrt{2}}(|H\rangle|V\rangle\pm|V\rangle|H\rangle).
\end{eqnarray}
States in Eq. (\ref{bell1}) can be regarded as the case of C-GHZ state in Eq. (\ref{logic}) with $N=M=2$.
\begin{figure}[!h]
\begin{center}
\includegraphics[width=9cm,angle=0]{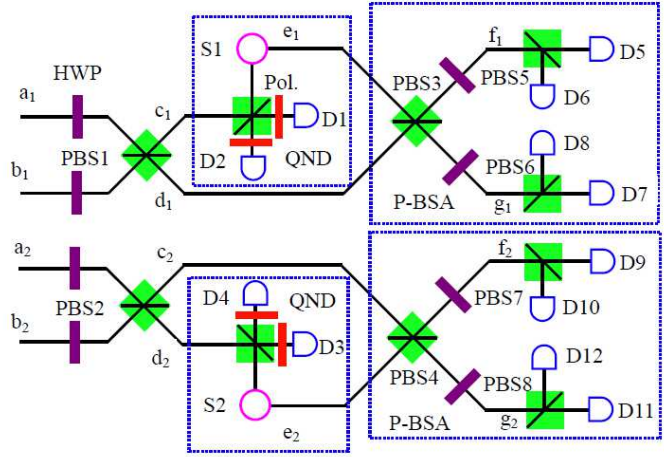}
\caption{ Protocol for logic Bell-state analysis. The QND is the teleportation-based probabilistic quantum nondemolition
measurement with an ancillary entangled photon pair, which is first experimentally demonstrated in the hyperentanglement Bell-state analysis \cite{bell14}. An incoming photon
can cause a coincidence detection after the beam splitter. Subsequently, it can herald its
presence and meanwhile can faithfully  teleport its arbitrary unknown quantum state
 to a free-flying photon for further application. The P-BSA is the polarization Bell-state analysis, which can completely distinguish
 $|\phi^{+}\rangle$ from $|\phi^{-}\rangle$. Pol. is the  linear polarizer.
}
\end{center}
\end{figure}
From Fig. 1, we first let four photons pass through four HWPs, respectively. The HWP can make
$|H\rangle\rightarrow \frac{1}{\sqrt{2}}(|H\rangle+|V\rangle)$, and $|V\rangle\rightarrow \frac{1}{\sqrt{2}}(|H\rangle-|V\rangle)$.
The HWPs will make the state $|\phi^{+}\rangle$ do not change, while $|\phi^{-}\rangle$ become $|\psi^{+}\rangle$. Therefore, after passing
through four HWPs, the four logic Bell states can evolve to
\begin{eqnarray}
|\Phi^{\pm}\rangle_{AB}\rightarrow\frac{1}{\sqrt{2}}(|\phi^{+}\rangle_{A}|\phi^{+}\rangle_{B}\pm|\psi^{+}\rangle_{A}|\psi^{+}\rangle_{B}),\nonumber\\
|\Psi^{\pm}\rangle_{AB}\rightarrow\frac{1}{\sqrt{2}}(|\phi^{+}\rangle_{A}|\psi^{+}\rangle_{B}\pm|\psi^{+}\rangle_{A}|\phi^{+}\rangle_{B}).\label{bell}
\end{eqnarray}
 States $|\Phi^{\pm}\rangle_{AB}$ can be written as
\begin{eqnarray}
&&|\Phi^{\pm}\rangle_{AB}=\frac{1}{\sqrt{2}}(|\phi^{+}\rangle_{A}|\phi^{+}\rangle_{B}\pm|\psi^{+}\rangle_{A}|\psi^{+}\rangle_{B})\nonumber\\
&=&\frac{1}{\sqrt{2}}[\frac{1}{\sqrt{2}}(|H\rangle_{a_{1}}|H\rangle_{a_{2}}+|V\rangle_{a_{1}}|V\rangle_{a_{2}})
\otimes\frac{1}{\sqrt{2}}(|H\rangle_{b_{1}}|H\rangle_{b_{2}}+|V\rangle_{b_{1}}|V\rangle_{b_{2}})\nonumber\\
&\pm&\frac{1}{\sqrt{2}}(|H\rangle_{a_{1}}|V\rangle_{a_{2}}+|V\rangle_{a_{1}}|H\rangle_{a_{2}})
\otimes\frac{1}{\sqrt{2}}(|H\rangle_{b_{1}}|V\rangle_{b_{2}}+|V\rangle_{b_{1}}|H\rangle_{b_{2}})]\nonumber\\
&=&\frac{1}{2\sqrt{2}}[(|H\rangle_{a_{1}}|H\rangle_{a_{2}}|H\rangle_{b_{1}}|H\rangle_{b_{2}}
+|H\rangle_{a_{1}}|H\rangle_{a_{2}}|V\rangle_{b_{1}}|V\rangle_{b_{2}}\nonumber\\
&+&|V\rangle_{a_{1}}|V\rangle_{a_{2}}|H\rangle_{b_{1}}|H\rangle_{b_{2}}
+|V\rangle_{a_{1}}|V\rangle_{a_{2}}|V\rangle_{b_{1}}|V\rangle_{b_{2}})\nonumber\\
&\pm&(|H\rangle_{a_{1}}|V\rangle_{a_{2}}|H\rangle_{b_{1}}|V\rangle_{b_{2}}+|H\rangle_{a_{1}}|V\rangle_{a_{2}}|V\rangle_{b_{1}}|H\rangle_{b_{2}}\nonumber\\
&+&|V\rangle_{a_{1}}|H\rangle_{a_{2}}|H\rangle_{b_{1}}|V\rangle_{b_{2}}+|V\rangle_{a_{1}}|H\rangle_{a_{2}}|V\rangle_{b_{1}}|H\rangle_{b_{2}})].\label{1}
\end{eqnarray}

States $|\Psi^{\pm}\rangle_{AB}$ can be written as
\begin{eqnarray}
&&|\Psi^{\pm}\rangle_{AB}=\frac{1}{\sqrt{2}}(|\phi^{+}\rangle_{A}|\psi^{+}\rangle_{B}\pm|\psi^{+}\rangle_{A}|\phi^{+}\rangle_{B})\nonumber\\
&=&\frac{1}{\sqrt{2}}[\frac{1}{\sqrt{2}}(|H\rangle_{a_{1}}|H\rangle_{a_{2}}+|V\rangle_{a_{1}}|V\rangle_{a_{2}})
\otimes\frac{1}{\sqrt{2}}(|H\rangle_{b_{1}}|V\rangle_{b_{2}}+|V\rangle_{b_{1}}|H\rangle_{b_{2}})\nonumber\\
&\pm&\frac{1}{\sqrt{2}}(|H\rangle_{a_{1}}|V\rangle_{a_{2}}+|V\rangle_{a_{1}}|H\rangle_{a_{2}})
\otimes\frac{1}{\sqrt{2}}(|H\rangle_{b_{1}}|H\rangle_{b_{2}}+|V\rangle_{b_{1}}|V\rangle_{b_{2}})]\nonumber\\
&=&\frac{1}{2\sqrt{2}}[(|H\rangle_{a_{1}}|H\rangle_{a_{2}}|H\rangle_{b_{1}}|V\rangle_{b_{2}}
+|H\rangle_{a_{1}}|H\rangle_{a_{2}}|V\rangle_{b_{1}}|H\rangle_{b_{2}}\nonumber\\
&+&|V\rangle_{a_{1}}|V\rangle_{a_{2}}|H\rangle_{b_{1}}|V\rangle_{b_{2}}
+|V\rangle_{a_{1}}|V\rangle_{a_{2}}|V\rangle_{b_{1}}|H\rangle_{b_{2}})\nonumber\\
&\pm&(|H\rangle_{a_{1}}|V\rangle_{a_{2}}|H\rangle_{b_{1}}|H\rangle_{b_{2}}+|H\rangle_{a_{1}}|V\rangle_{a_{2}}|V\rangle_{b_{1}}|V\rangle_{b_{2}}\nonumber\\
&+&|V\rangle_{a_{1}}|H\rangle_{a_{2}}|H\rangle_{b_{1}}|H\rangle_{b_{2}}+|V\rangle_{a_{1}}|H\rangle_{a_{2}}|V\rangle_{b_{1}}|V\rangle_{b_{2}})].\label{2}
\end{eqnarray}

Subsequently, we let four photons pass through the PBS1 and PBS2, respectively.
The PBS  can fully transmit the $|H\rangle$ polarized photon
and reflect the $|V\rangle$ polarized photon, respectively. By  selecting the cases where the spatial modes $c_{1}$, $d_{1}$, $c_{2}$ and $d_{2}$ all contain one photon, $|\Phi^{\pm}\rangle_{AB}$ will collapse to
\begin{eqnarray}
|\Phi^{\pm}\rangle_{AB}&\rightarrow&\frac{1}{2}[(|H\rangle_{c_{1}}|H\rangle_{c_{2}}|H\rangle_{d_{1}}|H\rangle_{d_{2}}
+|V\rangle_{c_{1}}|V\rangle_{c_{2}}|V\rangle_{d_{1}}|V\rangle_{d_{2}})\nonumber\\
&\pm&(|H\rangle_{c_{1}}|V\rangle_{c_{2}}|H\rangle_{d_{1}}|V\rangle_{d_{2}}+|V\rangle_{c_{1}}|H\rangle_{c_{2}}|V\rangle_{d_{1}}|H\rangle_{d_{2}})]\nonumber\\
&=&|\phi^{\pm}\rangle_{c_{1}d_{1}}\otimes|\phi^{\pm}\rangle_{c_{2}d_{2}}.\label{collapse1}
\end{eqnarray}
On the other hand, states $|\Psi^{\pm}\rangle_{AB}$ cannot make all the spatial modes $c_{1}$, $d_{1}$, $c_{2}$ and $d_{2}$ contain one photon.
For example, item $|H\rangle_{a_{1}}|H\rangle_{a_{2}}|V\rangle_{b_{1}}|H\rangle_{b_{2}}$ will make spatial mode $d_{1}$ contain two photons but spatial mode $c_{1}$ contain no photon. Item $|H\rangle_{a_{1}}|V\rangle_{a_{2}}|H\rangle_{b_{1}}|H\rangle_{b_{2}}$ will make spatial mode
$c_{2}$ contain two photons, but no photon in the spatial mode $d_{2}$.

In order to ensure all the four spatial modes contain one photon, our approach exploits quantum non-demolition (QND) measurement.
It means that a single photon can be observed without being destroyed, and
its quantum information  can be kept. Quantum teleportation is a powerful approach to implement the QND measurement. Adopting the quantum teleportation to implement the QND measurement for realizing the Bell state analysis was first discussed in Ref.\cite{bell14}. It will be detailed in Method Section.

After both successful teleportation, states $|\Phi^{\pm}\rangle_{AB}$ become $|\phi^{\pm}\rangle_{e_{1}d_{1}}\otimes|\phi^{\pm}\rangle_{e_{2}d_{2}}$, while states $|\Psi^{\pm}\rangle_{AB}$ never lead to both successful teleportation. States $|\phi^{\pm}\rangle$ can be easily distinguished with polarization Bell-state analysis (P-BSA) \cite{ghzanalysis1}, as shown in Fig. 1. Briefly speaking, we let the four photon pass through two PBSs and four HWPs for a second time, respectively.
After that, state $|\phi^{+}\rangle_{e_{1}d_{1}}\otimes|\phi^{+}\rangle_{e_{2}d_{2}}$ will not change, while state $|\phi^{-}\rangle_{e_{1}d_{1}}\otimes|\phi^{-}\rangle_{e_{2}d_{2}}$ will become $|\psi^{+}\rangle_{e_{1}d_{1}}\otimes|\psi^{+}\rangle_{e_{2}d_{2}}$.
 According to the coincidence measurement, we can finally distinguish the states
$|\Phi^{\pm}\rangle_{AB}$. For example, if the coincidence measurement result is one of  D5D7D9D11, D5D7D10D12, D6D8D9D11 or D6D8D10D12, the original state must be $|\Phi^{+}\rangle_{AB}$. On the other hand, if the coincidence measurement result is one of  D5D8D9D12,  D5D8D10D11, D6D7D9D12 or D6D7D10D11, it must be $|\Phi^{-}\rangle_{AB}$. In this way, we can completely distinguish the states $|\Phi^{\pm}\rangle_{AB}$.

 In this protocol, each logic qubit is encoded in a polarized Bell state. Actually, if the logic qubit is
encoded in a $M$-photon GHZ state, we can also discriminate two logic Bell states. The generalized four logic Bell states  can be described as
\begin{eqnarray}
|\Phi_{M}^{\pm}\rangle_{AB}&=&\frac{1}{\sqrt{2}}(|GHZ_{M}^{+}\rangle_{A}|GHZ_{M}^{+}\rangle_{B}
\pm|GHZ_{M}^{-}\rangle_{A}|GHZ_{M}^{-}\rangle_{B}),\nonumber\\
|\Psi_{M}^{\pm}\rangle_{AB}&=&\frac{1}{\sqrt{2}}(|GHZ_{M}^{+}\rangle_{A}|GHZ_{M}^{-}\rangle_{B}
\pm|GHZ_{M}^{-}\rangle_{A}|GHZ_{M}^{+}\rangle_{B}).\label{nlogicbell}
\end{eqnarray}

In order to explain this protocol clearly, we first let $M=3$ for simple. If $M=3$, the three-photon polarized GHZ states $|GHZ_{3}^{\pm}\rangle$ can be written as
\begin{eqnarray}
|GHZ_{3}^{\pm}\rangle=\frac{1}{\sqrt{2}}(|H\rangle|H\rangle|H\rangle\pm|V\rangle|V\rangle|V\rangle).
\end{eqnarray}
After performing the Hadamard operation on each photon, states $|\Phi_{3}^{\pm}\rangle_{AB}$ and $|\Psi_{3}^{\pm}\rangle_{AB}$ can be transformed to
\begin{eqnarray}
|\Phi_{3}^{\pm}\rangle_{AB}&=&\frac{1}{\sqrt{2}}(|GHZ_{3}^{+}\rangle^{\perp}_{A}|GHZ_{3}^{+}\rangle^{\perp}_{B}
\pm|GHZ_{3}^{-}\rangle^{\perp}_{A}|GHZ_{3}^{-}\rangle^{\perp}_{B}),\nonumber\\
|\Psi_{3}^{\pm}\rangle_{AB}&=&\frac{1}{\sqrt{2}}(|GHZ_{3}^{+}\rangle^{\perp}_{A}|GHZ_{3}^{-}\rangle^{\perp}_{B}
\pm|GHZ_{3}^{-}\rangle^{\perp}_{A}|GHZ_{3}^{+}\rangle^{\perp}_{B}).\label{nlogicbell1}
\end{eqnarray}
Here
\begin{eqnarray}
|GHZ_{3}^{+}\rangle^{\perp}&=&\frac{1}{2}(|H\rangle|H\rangle|H\rangle+|H\rangle|V\rangle|V\rangle
+|V\rangle|H\rangle|V\rangle+|V\rangle|V\rangle|H\rangle),\nonumber\\
|GHZ_{3}^{-}\rangle^{\perp}&=&\frac{1}{2}(|H\rangle|H\rangle|V\rangle+|H\rangle|V\rangle|H\rangle
+|V\rangle|H\rangle|H\rangle+|V\rangle|V\rangle|V\rangle).
\end{eqnarray}
From Eq. (\ref{nlogicbell1}), after performing the Hadamard operation, compared with the states in Eq. (\ref{nlogicbell}), states $|\Phi_{3}^{\pm}\rangle_{AB}$ and $|\Psi_{3}^{\pm}\rangle_{AB}$ have the different
form, for the logic qubit encoded in the GHZ state $|GHZ_{3}^{\pm}\rangle$  cannot be transformed to another GHZ state, which is quite different from the Bell states. States $|\Phi_{3}^{\pm}\rangle_{AB}$ can be rewritten as
\begin{eqnarray}
|\Phi_{3}^{\pm}\rangle_{AB}&=&\frac{1}{4\sqrt{2}}[(|H\rangle_{a_{1}}|H\rangle_{a_{2}}|H\rangle_{a_{3}}+|H\rangle_{a_{1}}|V\rangle_{a_{2}}|V\rangle_{a_{3}}\nonumber\\
&+&|V\rangle_{a_{1}}|H\rangle_{a_{2}}|V\rangle_{a_{3}}+|V\rangle_{a_{1}}|V\rangle_{a_{2}}|H\rangle_{a_{3}})\nonumber\\
&\otimes&(|H\rangle_{b_{1}}|H\rangle_{b_{2}}|H\rangle_{b_{3}}+|H\rangle_{b_{1}}|V\rangle_{b_{2}}|V\rangle_{b_{3}}\nonumber\\
&+&|V\rangle_{b_{1}}|H\rangle_{b_{2}}|V\rangle_{b_{3}}+|V\rangle_{b_{1}}|V\rangle_{b_{2}}|H\rangle_{b_{3}})\nonumber\\
&\pm&(|H\rangle_{a_{1}}|H\rangle_{a_{2}}|V\rangle_{a_{3}}+|H\rangle_{a_{1}}|V\rangle_{a_{2}}|H\rangle_{a_{3}}\nonumber\\
&+&|V\rangle_{a_{1}}|H\rangle_{a_{2}}|H\rangle_{a_{3}}+|V\rangle_{a_{1}}|V\rangle_{a_{2}}|V\rangle_{a_{3}})\nonumber\\
&\otimes&(|H\rangle_{b_{1}}|H\rangle_{b_{2}}|V\rangle_{b_{3}}+|H\rangle_{b_{1}}|V\rangle_{b_{2}}|H\rangle_{b_{3}}\nonumber\\
&+&|V\rangle_{b_{1}}|H\rangle_{b_{2}}|H\rangle_{b_{3}}+|V\rangle_{b_{1}}|V\rangle_{b_{2}}|V\rangle_{b_{3}})].\label{3}
\end{eqnarray}
From Fig. 1, if the logic qubit is three-photon polarized GHZ state, we should add the same setup in spatial modes $a_{3}$ and $b_{3}$, as it is in $a_{1}$ and $b_{1}$.
Certainly, we require three QNDs to complete the task.
  If we pick up the case that all the spatial modes $c_{1}$, $d_{1}$, $c_{2}$, $d_{2}$, $c_{3}$ and $d_{3}$ contain one photon,
states $|\Phi_{3}^{\pm}\rangle_{AB}$ will collapse to
\begin{eqnarray}
|\Phi_{3}^{\pm}\rangle_{AB}&\rightarrow&\frac{1}{2\sqrt{2}}[(|H\rangle_{a_{1}}|H\rangle_{a_{2}}|H\rangle_{a_{3}}|H\rangle_{b_{1}}|H\rangle_{b_{2}}|H\rangle_{b_{3}}\nonumber\\
&+&|H\rangle_{a_{1}}|V\rangle_{a_{2}}|V\rangle_{a_{3}}|H\rangle_{b_{1}}|V\rangle_{b_{2}}|V\rangle_{b_{3}}\nonumber\\
&+&|V\rangle_{a_{1}}|H\rangle_{a_{2}}|V\rangle_{a_{3}}|V\rangle_{b_{1}}|H\rangle_{b_{2}}|V\rangle_{b_{3}}\nonumber\\
&+&|V\rangle_{a_{1}}|V\rangle_{a_{2}}|H\rangle_{a_{3}}|V\rangle_{b_{1}}|V\rangle_{b_{2}}|H\rangle_{b_{3}})\nonumber\\
&\pm&(|H\rangle_{a_{1}}|H\rangle_{a_{2}}|V\rangle_{a_{3}}|H\rangle_{b_{1}}|H\rangle_{b_{2}}|V\rangle_{b_{3}}\nonumber\\
&+&|H\rangle_{a_{1}}|V\rangle_{a_{2}}|H\rangle_{a_{3}}|H\rangle_{b_{1}}|V\rangle_{b_{2}}|H\rangle_{b_{3}}\nonumber\\
&+&|V\rangle_{a_{1}}|H\rangle_{a_{2}}|H\rangle_{a_{3}}|V\rangle_{b_{1}}|H\rangle_{b_{2}}|H\rangle_{b_{3}}\nonumber\\
&+&|V\rangle_{a_{1}}|V\rangle_{a_{2}}|V\rangle_{a_{3}}|V\rangle_{b_{1}}|V\rangle_{b_{2}}|V\rangle_{b_{3}})]\nonumber\\
&=&|\phi^{\pm}\rangle_{a_{1}b_{1}}\otimes|\phi^{\pm}\rangle_{a_{2}b_{2}}\otimes|\phi^{\pm}\rangle_{a_{3}b_{3}}.\label{collapse2}
\end{eqnarray}
In order to complete such task, we require three pairs of polarized entangled states as auxiliary to perform the QND and
coincidence measurement. States $|\Psi_{3}^{\pm}\rangle_{AB}$ never lead to the case that all the  spatial modes
 $c_{1}$, $d_{1}$, $c_{2}$, $d_{2}$, $c_{3}$ and $d_{3}$ contain one photon, which can be excluded automatically. The next step is also to distinguish
 the state $|\phi^{+}\rangle$ from $|\phi^{-}\rangle$, which is analogy with the previous description. In this way, we can completely distinguish the state
 $|\Phi_{3}^{+}\rangle_{AB}$ from $|\Phi_{3}^{-}\rangle_{AB}$.

 Obviously, this approach can be extended to distinguish the logic Bell-state with the logic qubits  encoded in the $M$-photon GHZ state
 $|GHZ^{\pm}_{M}\rangle$, by adding the same setup in the spatial modes $a_{3}$ and $b_{3}$, $a_{4}$ and $b_{4}$, $\cdots$, and so on.  With the help of QNDs and coincidence measurement, we can pick up the cases where all the spatial modes $c_{1}$, $d_{1}$, $c_{2}$, $d_{2}$, $\cdots$, $c_{M}$ and $d_{M}$ exactly contain one photon, which make the states $|\Phi_{M}^{\pm}\rangle_{AB}$ collapse to
 $|\phi^{\pm}\rangle_{a_{1}b_{1}}\otimes|\phi^{\pm}\rangle_{a_{2}b_{2}}\cdots|\phi^{\pm}\rangle_{a_{M}b_{M}}$. Each state $|\phi^{\pm}\rangle$ can be distinguished by the P-BSA.  In this way, one can distinguish two logic Bell states with each logic qubit being the arbitrary $M$-photon GHZ state.

The GHZ state also plays an important role in fundamental
tests of quantum mechanics and it exhibits a conflict with local
realism for non-statistical predictions of quantum mechanics \cite{panrev}.
The first polarized GHZ state analysis was discussed by Pan and Zeilinger \cite{ghzanalysis1}. In their protocol, assisted with PBSs and HWPs, they can
conveniently identify two of the three-particle GHZ states. Interestingly, our protocol described above can also be extended to the C-GHZ state analysis. The C-GHZ state can be described as
\begin{eqnarray}
|\Phi^{\pm}_{1}\rangle_{N,2}&=&\frac{1}{\sqrt{2}}(|\phi^{+}\rangle^{\otimes N}\pm|\phi^{-}\rangle^{\otimes N}),\nonumber\\
|\Phi^{\pm}_{2}\rangle_{N,2}&=&\frac{1}{\sqrt{2}}(|\phi^{-}\rangle|\phi^{+}\rangle^{\otimes N-1}\pm|\phi^{+}\rangle|\phi^{-}\rangle^{\otimes N-1}),\nonumber\\
&\cdots&,\nonumber\\
|\Phi^{\pm}_{2^{N-1}}\rangle_{N,2}&=&\frac{1}{\sqrt{2}}(|\phi^{+}\rangle^{\otimes N-1}|\phi^{-}\rangle\pm|\phi^{-}\rangle^{\otimes N-1}|\phi^{+}\rangle.\label{multi2}
\end{eqnarray}
\begin{figure}[!h]
\begin{center}
\includegraphics[width=8cm,angle=0]{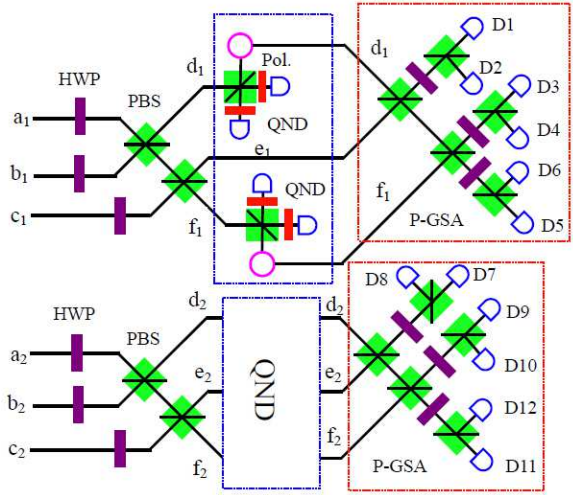}
\caption{Protocol for C-GHZ state analysis with $N=3$. The QND in the spatial modes $d_{2}$, $e_{2}$ and $f_{2}$ is the same as the QND in $d_{1}$, $e_{1}$ and $f_{1}$.
The P-GSA is the polarized GHZ-state analyzer, which was first described in Ref. \cite{ghzanalysis1}.}
\end{center}
\end{figure}
We let the logic qubits be the Bell states $|\phi^{\pm}\rangle$ and still take $N=3$ for example. From Fig. 2, after passing through the HWPs, the C-GHZ states can be described as
\begin{eqnarray}
&&|\Phi^{\pm}_{1}\rangle_{3,2}=\frac{1}{\sqrt{2}}(|\phi^{+}\rangle_{A}|\phi^{+}\rangle_{B}|\phi^{+}\rangle_{C}
\pm|\psi^{+}\rangle_{A}|\psi^{+}\rangle_{B}|\psi^{+}\rangle_{C}),\nonumber\\
&&|\Phi^{\pm}_{2}\rangle_{3,2}=\frac{1}{\sqrt{2}}(|\psi^{+}\rangle_{A}|\phi^{+}\rangle_{B}|\phi^{+}\rangle_{C}
\pm|\phi^{+}\rangle_{A}|\psi^{+}\rangle_{B}|\psi^{+}\rangle_{C}),\nonumber\\
&&|\Phi^{\pm}_{3}\rangle_{3,2}=\frac{1}{\sqrt{2}}(|\phi^{+}\rangle_{A}|\psi^{+}\rangle_{B}|\phi^{+}\rangle_{C}
\pm|\psi^{+}\rangle_{A}|\phi^{+}\rangle_{B}|\psi^{+}\rangle_{C}),\nonumber\\
&&|\Phi^{\pm}_{4}\rangle_{3,2}=\frac{1}{\sqrt{2}}(|\phi^{+}\rangle_{A}|\phi^{+}\rangle_{B}|\psi^{+}\rangle_{C}
\pm|\psi^{+}\rangle_{A}|\psi^{+}\rangle_{B}|\phi^{+}\rangle_{C}).\label{lghz}
\end{eqnarray}
We let the six photons pass through four PBSs, respectively. If we pick up the cases in which all the spatial modes
$d_{1}$, $e_{1}$, $f_{1}$, $d_{2}$, $e_{2}$ and  $f_{2}$ exactly contain one photon, states $|\Phi^{\pm}_{1}\rangle_{3,2}$
will become
\begin{eqnarray}
|\Phi^{\pm}_{1}\rangle_{3,2}&\rightarrow&\frac{1}{2}[(|H\rangle_{a_{1}}|H\rangle_{b_{1}}|H\rangle_{c_{1}}|H\rangle_{a_{2}}|H\rangle_{b_{2}}|H\rangle_{c_{2}}\nonumber\\
&+&|V\rangle_{a_{1}}|V\rangle_{b_{1}}|V\rangle_{c_{1}}|V\rangle_{a_{2}}|V\rangle_{b_{2}}|V\rangle_{c_{2}})\nonumber\\
&\pm&(|H\rangle_{a_{1}}|H\rangle_{b_{1}}|H\rangle_{c_{1}}|V\rangle_{a_{2}}|V\rangle_{b_{2}}|V\rangle_{c_{2}})\nonumber\\
&+&|V\rangle_{a_{1}}|V\rangle_{b_{1}}|V\rangle_{c_{1}}|H\rangle_{a_{2}}|H\rangle_{b_{2}}|H\rangle_{c_{2}})]\nonumber\\
&=&\frac{1}{\sqrt{2}}(|H\rangle_{a_{1}}|H\rangle_{b_{1}}|H\rangle_{c_{1}}\pm|V\rangle_{a_{1}}|V\rangle_{b_{1}}|V\rangle_{c_{1}})\nonumber\\
&\otimes&\frac{1}{\sqrt{2}}(|H\rangle_{a_{2}}|H\rangle_{b_{2}}|H\rangle_{c_{2}}\pm|V\rangle_{a_{2}}|V\rangle_{b_{2}}|V\rangle_{c_{2}})\nonumber\\
&=&|GHZ_{3}^{\pm}\rangle_{a_{1}b_{1}c_{1}}\otimes|GHZ_{3}^{\pm}\rangle_{a_{2}b_{2}c_{2}}.\label{lghz1}
\end{eqnarray}
In order to complete this task, we also exploit the QNDs. As shown in Fig. 2, we require four  QNDs, which are the same as those in Fig. 1. The QNDs in spatial
modes  $d_{2}$, $e_{2}$ and  $f_{2}$ is the same as those in the spatial modes $d_{1}$, $e_{1}$ and $f_{1}$. From Eq. (\ref{lghz1}), if all the spatial modes $d_{1}$, $e_{1}$, $f_{1}$, $d_{2}$, $e_{2}$ and  $f_{2}$ exactly contain one photon, the initial states $|\Phi^{\pm}_{1}\rangle_{3,2}$ will collapse to the standard polarized
GHZ states $|GHZ_{3}^{\pm}\rangle_{a_{1}b_{1}c_{1}}\otimes|GHZ_{3}^{\pm}\rangle_{a_{2}b_{2}c_{2}}$.  States $|GHZ_{3}^{\pm}\rangle$ can be deterministically
distinguished by the setup of polarized GHZ-state analysis (P-GSA), as shown in Fig. 2. The P-GSA was first described in Ref.\cite{ghzanalysis1}. Briefly speaking, $|GHZ_{3}^{+}\rangle_{a_{1}b_{1}c_{1}}$ leads to coincidence between detectors D1D3D5, D1D4D6, D2D3D6 or D2D4D5, and $|GHZ_{3}^{-}\rangle_{a_{1}b_{1}c_{1}}$ leads to coincidence between detectors D2D4D6, D1D4D5, D2D3D5 or D1D3D6. State $|GHZ_{3}^{\pm}\rangle_{a_{2}b_{2}c_{2}}$ can be distinguished in the same principle.  In this way, we can distinguish two states $|\Phi^{\pm}_{1}\rangle_{3,2}$
from the eight states as described in Eq. (\ref{lghz}).

\begin{figure}[!h]
\begin{center}
\includegraphics[width=8cm,angle=0]{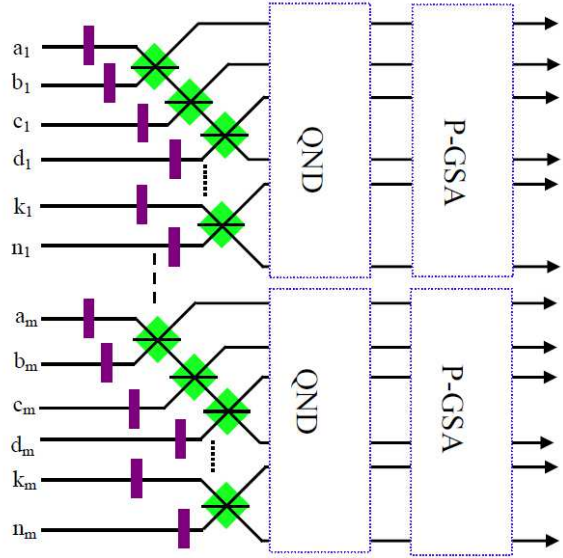}
\caption{Protocol for C-GHZ state analysis with arbitrary $N$ and $M$. The QNDs are used to ensure that each spatial mode contains one photon, which can project
the original state to one of the $N$-photon polarized GHZ states $|GHZ_{N}^{\pm}\rangle$. The P-GSA can distinguish $|GHZ_{N}^{\pm}\rangle$ \cite{ghzanalysis1}.}
\end{center}
\end{figure}

For the $N$-logic qubit C-GHZ state analysis, this protocol can also work. As shown in Fig. 3, if each logic qubit is a Bell state, we let the photons in spatial modes $a_{1}$, $b_{1}$, $\cdots$, $n_{1}$ and  $a_{2}$, $b_{2}$, $\cdots$, $n_{2}$ pass through the $N-1$ PBS, respectively. By using  QNDs to ensure each of the spatial modes behind the $N-1$ PBSs contains one photon, it will project the states $|\Phi^{\pm}_{1}\rangle_{N,2}$ to $|GHZ_{N}^{\pm}\rangle_{a_{1}b_{1}\cdots n_{1}}\otimes|GHZ_{N}^{\pm}\rangle_{a_{2}b_{2}\cdots n_{2}}$, which can be completely distinguished by P-GSA as described in Ref.\cite{ghzanalysis1}. We can also distinguish two C-GHZ states with arbitrary $N$ and $M$.
By adding the same setup in the spatial modes $a_{3}$, $b_{3}$, $\cdots$, $n_{3}$, $\cdots$, $a_{m}$, $b_{m}$, $\cdots$, $n_{m}$, we can project the C-GHZ states
to $|\Phi^{\pm}_{1}\rangle_{N,M}$ to $|GHZ_{N}^{\pm}\rangle_{a_{1}b_{1}\cdots n_{1}}\otimes|GHZ_{N}^{\pm}\rangle_{a_{2}b_{2}\cdots n_{2}}\otimes\cdots\otimes|GHZ_{N}^{\pm}\rangle_{a_{m}b_{m}\cdots n_{m}}$,  with the help of QNDs. Each pair of $N$-photon polarization GHZ states $|GHZ_{N}^{\pm}\rangle$ can be well distinguished. In this way, we can identify $|\Phi^{\pm}_{1}\rangle_{N,M}$ from arbitrary C-GHZ state completely.

\section*{Discussion}
So far, we have completely described our logic Bell-state and C-GHZ state analysis. In the logic Bell-state analysis, we can completely distinguish the states
$|\Phi^{\pm}\rangle$ from the four logic Bell states. For arbitrary C-GHZ state analysis, we can also distinguish two states $|\Phi^{\pm}_{1}\rangle_{N,M}$ from the arbitrary $N$-logic-qubit C-GHZ states.
It is interesting to discuss the possible experiment realization.   In a practical experiment, one challenge comes from the multi-photon entanglement, for we require two polarization Bell states as auxiliary and the whole protocol requires eight photons totally.
Fortunately, the eight-photon entanglement has been observed with cascaded entanglement sources
 \cite{eightphoton1,eightphoton2}. The other challenge is the  QND with linear optics \cite{linearQND1,linearQND2}. From Fig. 2, the QND exploits Hong-Ou-Mandel interference \cite{hongou} between two undistinguishable photons with good spatial, time and spectral. As shown in Ref.\cite{bell14}, the Hong-Ou Mandel interference of multiple independent photons has been well observed with the visibility is $0.73\pm0.03$. Different from Ref.\cite{bell14}, we are required to prepare two independent pairs of entangled photons at the same time. This challenge can also be overcome with  cascaded entanglement sources, which can synchronized generate two pairs of polarized entangled photons.  This approach has also been realized in previous experimental quantum teleportation of a two-qubit composite system \cite{twoqnd}. The final verification of the Bell-state analysis relies on the coincidence
detection counts of the eight photons, with four photons coming from the QNDs and four coming from the P-BSA.  This   technical challenge of very low eight photon coincidence count rate was also overcome in the previous experiment by using brightness of entangled photons \cite{eightphoton1,eightphoton2}. Finally, let us briefly discuss the total success probability of this protocol. In a practical experiment, we should both consider the efficiency of the entanglement source and single-photon detector. Usually, we exploit the spontaneous parametric down-conversion (SPDC) source to implement the entanglement source \cite{double}. In order to distinguish C-GHZ state with $M$ and $N$, we require $(M-1)N$ entanglement sources and $[2(M-1)+M]N$ single-photon detectors. Suppose that the efficiency of the SPDC source is $p_{s}$. A practical single-photon detector can be regarded as a perfect detection with a loss element in front of it. The probability of detecting a photon can then be given as $p_{d}$. Therefor, the total success probability $P_{t}$ can be written as
\begin{eqnarray}
P_{t}=p^{(M-1)N}_{s}p^{[2(M-1)+M]N}_{d}.
\end{eqnarray}
As point out in Ref.\cite{bell14}, the mean numbers of photon pairs generated per pulse as $p_{s}\sim0.1$.  We let high-efficiency single-photon detectors with $p_{d}=0.9$. We calculate the total success probability $P_{t}$ altered with the $M$ and $N$. If $M=N=2$,  we can obtain $P_{t}\approx 0.00656$. In Fig. 4, the success probability is quite low, if $M$ increases.  From calculation, the imperfect entanglement source will greatly limit the total success probability.
This problem can in principle be eliminated in future by various methods,
such as  deterministic entangled photons \cite{entanglementsource}.

\begin{figure}[!h]
\begin{center}
\includegraphics[width=8cm,angle=0]{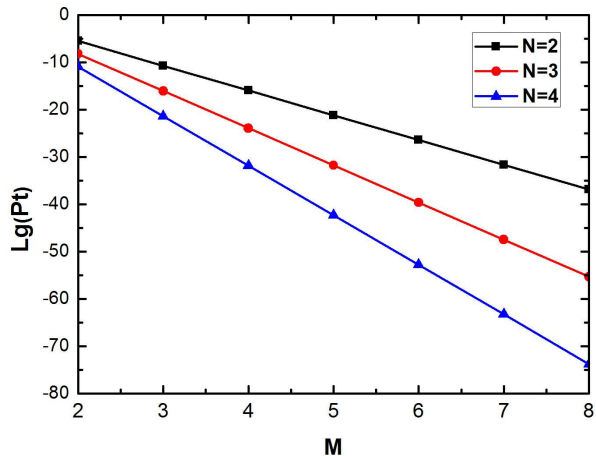}
\caption{Schematic of the success probability altered with the physical qubit number $M$. Here we let $N=2,3$ and 4, respectively.}
\end{center}
\end{figure}
In conclusion, we have proposed a feasible logic Bell-state analysis protocol. By exploiting the approach of teleportation-based QND, we can completely distinguish
two logic Bell states $|\Phi^{\pm}\rangle$ among four logic Bell-states. This protocol can also be extended to distinguish arbitrary C-GHZ state. We can also identify two C-GHZ states among $2^{N}$ C-GHZ states. The biggest advantage of this protocol is that it is based on the linear optics, so that it is feasible in current experimental technology. As the Bell-state analysis plays a key role in quantum communication, this protocol may provide an important application in large-scale fibre-based quantum networks and the quantum communication based on the logic qubit entanglement. Moreover,
this protocol may also be
useful for linear-optical quantum computation protocols whose building blocks are GHZ-type states.

\section*{Methods}
The QND is the key element in this protocol. Here we exploit the quantum teleportation to realize the QND. As shown in Fig. 1, both the entanglement sources $S1$ and $S2$ create a pair of polarized entangled state $|\phi^{+}\rangle$, respectively.
 If the spatial mode $c_{1}$ only contains a photon, a two-photon coincidence behind the PBS can occur with 50\% success probability to trigger
 a Bell-state analysis. Meanwhile, both single-photon detectors $D1$ and $D2$ register a photon also means that we can identify $|\phi^{+}\rangle$ with the success probability of $1/4$,
 which  is a successful teleportation. It can
 teleport the incoming photon in the spatial mode $c_{1}$ to a freely propagating photon in the spatial mode $e_{1}$.
 On the other hand, if the spatial mode $c_{1}$ contains no photon, the two-photon coincidence behind the PBS
cannot occur. We can notice the case and ignore the outgoing
photon.   Using a QND in one of the arm of
the PBS is sufficient. That is because the conserved total number of
eventually registered photons for the case of two photon in spatial mode $c_{1}$ or $d_{2}$ can be eliminated automatically by the final
coincidence measurement. In our protocol, the setup of teleportation can only distinguish one Bell state among the four with the success probability of the QND being 1/4. In this way, the total success probability of this protocol
is $1/4\times1/4\times1/2=1/32$.  By introducing a more complicated setup of teleportation which can distinguish two
polarized Bell states among the four \cite{bellstateanalysis2}, the success probability can be improved to $1/2\times1/2\times1/2=1/8$ in principle.

\section*{Acknowledgements}
This work is supported by the National Natural Science Foundation of China (Grant Nos. 11474168 and 61401222), the Natural Science Foundation of
Jiangsu under Grant No. BK20151502, the Qing Lan Project in Jiangsu Province, and the Priority Academic Development Program of Jiangsu Higher Education Institutions, China.

\section*{Author contributions}
Y. B. S. presented the idea,  L. Z. wrote the main manuscript text and  prepared figures 1-4. Both authors reviewed the manuscript.

\section*{Additional information}
Competing financial interests: The authors declare no competing financial interests.
\end{document}